\documentclass[prb,twocolumn,showpacs]{revtex4}
\usepackage{graphicx} 
\usepackage{amsmath,amsthm,amssymb,mathrsfs}
\usepackage{bm}

\begin{document}

\title{Relaxation time and critical slowing down of a spin-torque oscillator}

\author{Tomohiro Taniguchi${}^{1}$, Takahiro Ito${}^{2}$, Sumito Tsunegi${}^{1}$, Hitoshi Kubota${}^{1}$, and Yasuhiro Utsumi${}^{2}$}
 \affiliation{
${}^{1}$National Institute of Advanced Industrial Science and Technology (AIST), Spintronics Research Center, Tsukuba 305-8568, Japan \\
${}^{2}$Faculty of Engineering, Mie University, Tsu, Mie, 514-8507, Japan 
 }

 \begin{abstract}
{ 
The relaxation phenomena of spin-torque oscillators consisting of nanostructured ferromagnets are interesting research targets in magnetism. 
A theoretical study on the relaxation time of a spin-torque oscillator from one self-oscillation state to another is investigated. 
By solving the Landau-Lifshitz-Gilbert equation both analytically and numerically, 
it is shown that the oscillator relaxes to the self-oscillation state exponentially within a few nanoseconds, 
except when magnetization is close to a critical point. 
The relaxation rate, which is an inverse of relaxation time, is proportional to the current. 
On the other hand, a critical slowing down appears near the critical point, 
where relaxation is inversely proportional to time, 
and the relaxation time becomes on the order of hundreds of nanoseconds. 
These conclusions are primarily obtained for a spin-torque oscillator consisting of 
a perpendicularly magnetized free layer and an in-plane magnetized pinned layer, 
and are further developed for application to arbitrary types of spin-torque oscillators. 
}
 \end{abstract}

 \pacs{05.70.Jk, 75.78.Jp, 05.45.-a, 85.75.-d}
 \maketitle




\section{Introduction}
\label{sec:Introduction}

Limit cycles of magnetization with an oscillation frequency on the order of gigahertz may appear 
in nanostructured ferromagnetic/nonmagnetic multilayers as a result of injecting spin current 
[\onlinecite{kiselev03,rippard04,krivorotov05,houssameddine07,persson07,slavin09nature,rippard10,urazhdin10,sinha11,zeng12,suto12,rippard13,kubota13,tamaru14,kudo15,tsunegi16,tsunegi16APL,awad17}]. 
spin-torque oscillators utilizing this self-oscillation provide interesting phenomena in the field of nonlinear science such as synchronization. 
The spin-torque oscillator has also attracted much attention from the viewpoint of practical applications 
because its small size, compatibility with current technology, and unnecessity of resonators are great advantages 
for magnetic sensors, microwave generators, and neuromorphic architectures [\onlinecite{locatelli14,grollier16,kudo17}]. 
Considerable efforts have been put into the development of high-performance spin-torque oscillators. 
High emission power ($>10 \mu$W), high quality factor ($>10^{3}$), and wide frequency tunability ($>3$ GHz) 
have been achieved in several kinds of spin-torque oscillators 
through material investigations, structural improvements, and/or utilizing synchronization. 
These steady-state properties have also been well studied theoretically, 
using nonlinear auto-oscillator models and numerical simulations [\onlinecite{ebels08,bertotti09,slavin09,silva10,nakada12,khalsa15}].

The next critical issue is to clarify the transient phenomenon in the spin-torque oscillators. 
A rapid response to external forces is a highly desirable property because it determines the speed of devices. 
For example, the spin-torque oscillators show the transition from a self-oscillation state to another 
under the application of magnetic pulses, resulting in a frequency transition [\onlinecite{kudo10,suto11,nagasawa11}]. 
To use such a transition as the operating principle of magnetic sensors, 
the transition time should be less than nanosecond order. 
The transition time will be estimated by calculating the relaxation time to the final state. 
The relaxation phenomenon in spin-torque oscillators, however, has not yet been fully clarified, 
particularly from a theoretical point of view, 
despite several reports on experiments [\onlinecite{suto11,nagasawa11}] and numerical simulations [\onlinecite{kudo10,purbawati16}]. 
A full understanding of the relaxation phenomena in spin-torque oscillators 
is therefore highly desirable for further development in practical devices. 


In this paper, we investigate the relaxation time of a spin-torque oscillator theoretically. 
Analytical formulas describing the relaxation of the magnetization to the self-oscillation state are derived, 
based on the Landau-Lifshitz-Gilbert (LLG) equation. 
The relaxation occurs exponentially within a time scale on the order of nanoseconds, 
except when the spin-torque oscillator is close to a critical point. 
The validity of the analytical formula is confirmed by comparison with numerical simulations, 
verifying the fast relaxation of magnetization. 
On the other hand, a critical slowing down appears near the critical point, 
where a linear approximation to the LLG equation is no longer applicable. 
The relaxation near the critical point is described by algebraic functions, 
rather than exponentials, and is on the order of hundreds of nanoseconds. 
These conclusions are primarily obtained for a particular type of oscillator 
and then are further extended to search for arbitrary systems. 
The results provide a comprehensive description of 
the relaxation and critical phenomena in the spin-torque oscillators. 


This paper is organized as follows. 
In Sec. \ref{sec:Relaxation time in spin-torque oscillator with perpendicularly magnetized free layer}, 
the relaxation time in a spin-torque oscillator is studied analytically. 
We focus on a spin-torque oscillator consisting of a perpendicularly magnetized free layer and an in-plane magnetized pinned layer as an example. 
We also perform a comparison with numerical simulation. 
In Sec. \ref{sec:Generalization of theory}, we generalize a theory of the relaxation phenomenon in spin-torque oscillators 
and show that the exponential relaxation and critical slowing down appear in general cases. 
Section \ref{sec:Conclusion} shows the conclusions.


\section{Relaxation time in spin-torque oscillator with perpendicularly magnetized free layer}
\label{sec:Relaxation time in spin-torque oscillator with perpendicularly magnetized free layer}

In this section, we investigate the relaxation time of the spin-torque oscillator both analytically and numerically. 
The spin-torque oscillator in this section consists of a perpendicularly magnetized free layer 
and an in-plane magnetized pinned layer. 


\subsection{System description}
\label{sec:System description}



\begin{figure}
\centerline{\includegraphics[width=1.0\columnwidth]{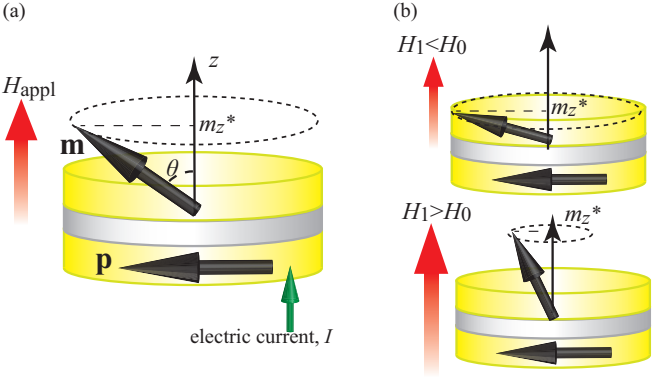}}
\caption{
         (a) Schematic view of the system. 
             In the self-oscillation state, the magnetization precesses on an orbit with a constant $m_{z}=m_{z}^{*}=\cos\theta$. 
         (b) When the magnitude of the external field changes from the initial value $H_{0}$ to a different value $H_{1}$, 
             the magnetization moves to a different self-oscillation state.
             When $H_{1}<H_{0}$, $m_{z}^{*}$ in the new state is smaller than the initial value, 
             whereas $m_{z}^{*}$ is larger than the initial value when $H_{1}>H_{0}$. 
         \vspace{-3ex}}
\label{fig:fig1}
\end{figure}



A schematic of the system under consideration is shown in Fig. \ref{fig:fig1}(a), 
where two ferromagnets sandwich a thin nonmagnet. 
The top and bottom ferromagnets correspond to the perpendicularly magnetized free and in-plane magnetized pinned layers, respectively [\onlinecite{kubota13}]. 
The unit vectors pointing in the magnetization direction of the free and pinned layers are denoted as 
$\mathbf{m}$ and $\mathbf{p}$, respectively. 
The $z$ axis is normal to the film plane, whereas the $x$ axis is parallel to the magnetization of the pinned layer, i.e., $\mathbf{p}=+\mathbf{e}_{x}$. 
The electric current $I$ is applied along the $z$ direction, which excites the magnetization dynamics by the spin-transfer effect [\onlinecite{slonczewski96,berger96}]. 
The positive current corresponds to the electron flow from the free to the pinned layer; 
i.e., the spin torque excited by the positive current prefers the antiparallel alignment of the magnetization. 
Recent experiments have shown that the magnetization in this type of spin-torque oscillator is 
well described by the LLG equation with the macrospin model [\onlinecite{rippard10,kubota13}], 
\begin{equation}
  \frac{d \mathbf{m}}{dt}
  =
  -\gamma
  \mathbf{m}
  \times
  \mathbf{H}
  -
  \gamma
  H_{\rm s}
  \mathbf{m}
  \times
  \left(
    \mathbf{p}
    \times
    \mathbf{m}
  \right)
  +
  \alpha
  \mathbf{m}
  \times
  \frac{d \mathbf{m}}{dt},
  \label{eq:LLG}
\end{equation}
where $\gamma$ and $\alpha$ are the gyromagnetic ratio and the Gilbert damping constant, respectively. 
The magnetic field $\mathbf{H}$ consists of the perpendicular anisotropy field and the external magnetic field $H_{\rm appl}$, expressed as 
\begin{equation}
  \mathbf{H}
  =
  \left[
    H_{\rm appl}
    +
    \left(
      H_{\rm K}
      -
      4\pi M
    \right)
    m_{z}
  \right]
  \mathbf{e}_{z},
  \label{eq:field}
\end{equation}
where $H_{\rm K}$ and $4\pi M$ are 
the crystalline and shape anisotropy fields, respectively. 
The magnetization has two energetically stable states at $m_{z}=\pm 1$. 
For convention, we assume that 
the magnetization maintains the stable state in the positive $z$ direction in the absence of the current. 
The spin-torque strength $H_{\rm s}$ [\onlinecite{slonczewski05}] is 
\begin{equation}
  H_{\rm s}
  =
  \frac{\hbar \eta I}{2e (1 + \lambda \mathbf{m}\cdot\mathbf{p}) MV}, 
  \label{eq:H_s}
\end{equation}
where $M$ and $V$ are the saturation magnetization and volume of the free layer, respectively. 
The spin polarization of the electric current and spin-torque asymmetry are denoted as $\eta$ and $\lambda$, respectively. 
The values of the parameters used in the following calculations are derived from Refs. [\onlinecite{kubota13,taniguchi13,tsunegi14}] as 
$M=1448.3$ emu/c.c., $H_{\rm K}=18.616$ kOe, $V=\pi \times 60^{2} \times 2$ nm${}^{3}$, 
$\eta=0.537$, $\lambda=0.288$, $\gamma = 1.764 \times 10^{7}$ rad/(Oe s), and $\alpha=0.005$. 




\begin{figure}
\centerline{\includegraphics[width=1.0\columnwidth]{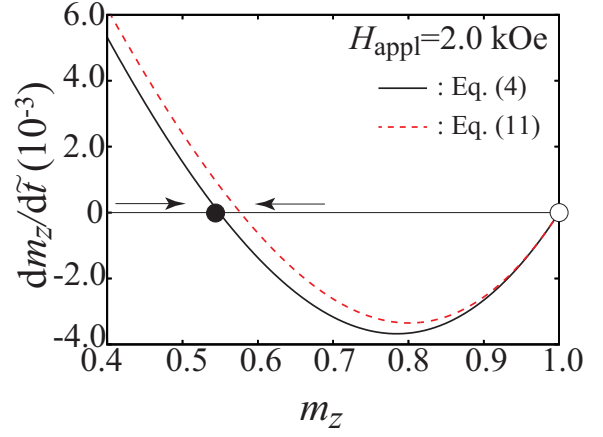}}
\caption{
         The time derivative of $m_{z}$, $d m_{z}/d \tilde{t}$, obtained from Eqs. (\ref{eq:LLG_averaged}) (black solid) and (\ref{eq:LLG_approx_1}) (red dotted), respectively,
         for the case with $I=2.5$ mA and $H_{\rm appl}=2.0$ kOe. 
         The black and white circles indicate the stable and unstable fixed points, respectively, 
         whereas the direction of the black arrows indicates the direction of the vector field, $dm_{z}/d \tilde{t}$. 
         \vspace{-3ex}}
\label{fig:fig2}
\end{figure}



A self-oscillation is excited when the spin torque balances with the damping torque during a precession, 
and the field torque, $-\gamma \mathbf{m} \times \mathbf{H}$ in Eq. (\ref{eq:LLG}), becomes 
the dominant term determining the magnetization dynamics. 
The field torque describes the steady precession of the magnetization on a trajectory with a constant cone angle $\theta=\cos^{-1}m_{z}$. 
Therefore, we use an approximation to average the LLG equation over the trajectories on a constant cone angle [\onlinecite{bertotti09,slavin09,dykman12}]. 
The LLG equation for $m_{z}$ is then given by 
\begin{equation}
\begin{split}
  \frac{d m_{z}}{d \tilde{t}}
  =&
  \alpha
  \left(
    m_{z}
    +
    h
  \right)
  \left(
    1
    -
    m_{z}^{2}
  \right)
\\
  &-
  \frac{h_{\rm s}}{\lambda}
  \left[
    \frac{1}{\sqrt{1 - \lambda^{2} (1-m_{z}^{2})}}
    -
    1
  \right]
  m_{z},
  \label{eq:LLG_averaged}
\end{split}
\end{equation}
where we introduce the following dimensionless quantities, for simplicity, 
\begin{equation}
  \tilde{t}
  \equiv
  \gamma
  \left(
    H_{\rm K}
    -
    4 \pi M 
  \right)
  t, 
\end{equation}
\begin{equation}
  h
  \equiv
  \frac{H_{\rm appl}}{H_{\rm K}-4\pi M},
\end{equation}
\begin{equation}
  h_{\rm s}
  \equiv
  \frac{\hbar \eta I}{2eMV(H_{\rm K}-4\pi M)}.
\end{equation}
The black solid line in Fig. \ref{fig:fig2} is an example of Eq. (\ref{eq:LLG_averaged}), 
showing $d m_{z}/d \tilde{t}$ as a function of $m_{z}$, 
where $I=2.5$ mA and $H_{\rm appl}=2.0$ kOe. 
There are two points satisfying $d m_{z}/d \tilde{t}=0$, which are called the fixed points [\onlinecite{strogatz01}]. 
The black arrows indicate the direction of $dm_{z}/d \tilde{t}$; 
i.e., the arrow points to the positive (negative) $m_{z}$ direction 
when $dm_{z}/d \tilde{t}$ is positive (negative). 
The fixed point at $m_{z}=+1$ (white circle) corresponds to an unstable fixed point, 
whereas the other fixed point at $m_{z}\simeq 0.55$ (black circle) is called a stable fixed point or attractor [\onlinecite{strogatz01}]. 
In the following, we denote the stable fixed point as $m_{z}^{*}$.  
The stable fixed point corresponds to the self-oscillation state. 
Therefore, the relaxation time can be defined as the time necessary to move 
from a certain $m_{z}$ to the stable fixed point. 
Equation (\ref{eq:LLG_averaged}) should be solved with respect to $m_{z}$ to evaluate the relaxation time. 
However, this equation is still difficult to solve. 
Thus, we use the two approximations shown below, 
i.e., an expansion of Eq. (\ref{eq:LLG_averaged}) around $|\lambda|=0$, Eq. (\ref{eq:LLG_approx_1}), or $m_{z} \simeq 1$, Eq. (\ref{eq:LLG_approx_3}). 
First, however, we discuss the validity of the approximation used in Eq. (\ref{eq:LLG_averaged}). 


\subsection{Validity of the averaging technique}
\label{sec:Validity of the averaging technique}

In the previous section, we applied the averaging technique of the LLG equation over a trajectory with a constant cone angle. 
This technique enables us to easily understand the magnetization dynamics analytically. 
In this section, we discuss both the applicability and limitations of this averaging technique. 


First, we explain the details of the approximation used in the averaging technique, 
which was briefly described in our previous work [\onlinecite{taniguchi13}]. 
The spin and damping torques should cancel each other to sustain the oscillation excited by the field torque, $-\gamma \mathbf{m} \times \mathbf{H}$. 
In the present system, however, it is impossible to balance these torques during the entire precession, 
because the torques have different angular dependences. 
This can be understood as follows. 
The damping torque, $-\alpha \gamma \mathbf{m}\times (\mathbf{m} \times \mathbf{H})$, points to the $z$ direction 
because the states $\mathbf{m}=\pm\mathbf{e}_{z}$ are energetically stable. 
On the other hand, the spin torque with the positive current forces the magnetization in the antiparallel direction to $\mathbf{p}=+\mathbf{e}_{x}$. 
The spin torque thus has a component parallel to the damping torque when $m_{x}>0$, 
whereas it has a component antiparallel to the damping torque when $m_{x}<0$. 
As a result, a complete cancellation between the spin and damping torques during the entire precession is impossible. 
We therefore need to relax the conditions to sustain the self-oscillation. 


For the typical ferromagnets, such as Co, Fe, Ni, and their composites, used in spin-torque oscillators, 
the damping constant $\alpha$ is on the order of $0.001-0.01$ [\onlinecite{oogane06}]. 
Therefore, the strength of the damping torque is at least two orders of magnitude smaller than that of the field torque. 
The strength of the spin torque is also smaller than the field torque, because it should compensate for the damping torque. 
Therefore, the difference between the exact trajectory of the magnetization dynamics and the trajectory determined by the field torque is small. 
Accordingly, it is a good approximation to average the LLG equation on a trajectory determined by the field torque, 
which in the present system corresponds to an orbit with a constant cone angle of $\theta=\cos^{-1}m_{z}$. 
The relaxed condition necessary to sustain the self-oscillation becomes such that the averaged spin and damping torques cancel each other. 
In other words, $d m_{z}/dt$ averaged over a constant cone angle is zero, as mentioned in Sec. \ref{sec:System description}. 




\begin{figure}
\centerline{\includegraphics[width=1.0\columnwidth]{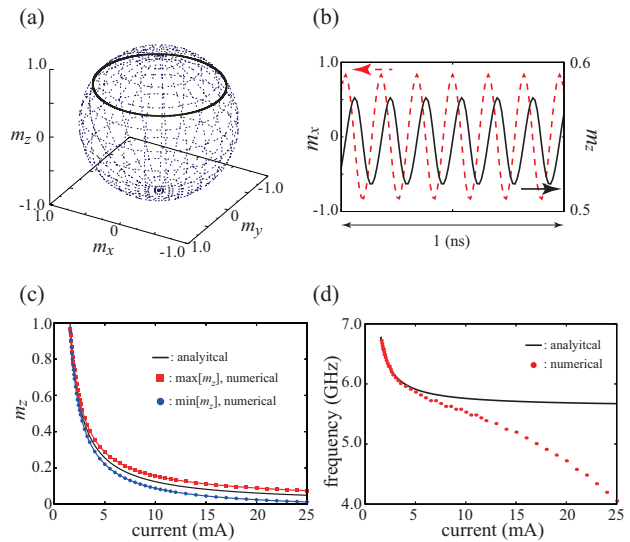}}
\caption{
         (a) A steady precession trajectory of the magnetization for $H_{\rm appl}=2.0$ kOe and $I=2.5$ mA. 
         (b) Time evolutions of $m_{x}$ (red dashed) and $m_{z}$ (black solid), respectively. 
         (c) Current dependences of the analytical $m_{z}$ (black line) and numerically evaluated ${\rm max}[m_{z}]$ (red square), and ${\rm min}[m_{z}]$ (blue circle). 
         (d) Current dependences of the oscillation frequencies estimated from the analytical theory (black line) and numerical simulations (red circle). 
         \vspace{-3ex}}
\label{fig:fig3}
\end{figure}



However, it is important to investigate the applicability of this averaging technique to validate the calculations in the following sections. 
Due to the angular dependence of the spin torque mentioned above, the exact solution of $m_{z}=\cos\theta$ is not a constant with time variance. 
We note that the difference of the exact $m_{z}$ from a constant value should be periodic, due to the periodicity of the self-oscillation. 
The averaging technique cannot take into account such an oscillating component. 
Therefore, we study the comparison between the exact and analytical solutions of the LLG equation for the present system. 
Figure \ref{fig:fig3}(a) shows the trajectory of the self-oscillation for $H_{\rm appl}=2.0$ kOe and $I=2.5$ mA, 
obtained by solving Eq. (\ref{eq:LLG}) numerically. 
The time evolutions of $m_{x}$ and $m_{z}$ are shown in Fig. \ref{fig:fig3}(b) by the red dashed and black solid lines, respectively. 
The important point in Fig. \ref{fig:fig3}(b) with respect to the discussion in this section is that
it provides clear evidence that the exact solution of $m_{z}$ is not a constant. 


We compare the constant $m_{z}$ in the analytical theory and oscillating $m_{z}$ as a function of the electric current. 
This is because the relation between the applied current $I$ (or voltage) and the frequency $f$ has been investigated in experiments [\onlinecite{rippard10,kubota13,tsunegi14}], 
and $m_{z}$ is related to the frequency. 
Using Eq. (\ref{eq:LLG_averaged}) with the condition $dm_{z}/dt=0$, 
the current necessary to excite self-oscillation with a constant cone angle $\theta=\cos^{-1}m_{z}$ is given by [\onlinecite{taniguchi13}] 
\begin{equation}
\begin{split}
  I(\theta)
  =&
  \frac{2 \alpha e \lambda MV}{\hbar \eta \cos\theta}
  \left(
    \frac{1}{\sqrt{1-\lambda^{2} \sin^{2}\theta}}
    -
    1
  \right)^{-1}
\\
  & 
  \times 
  \left[
    H_{\rm appl}
    +
    \left(
      H_{\rm K}
      -
      4\pi M 
    \right)
    \cos\theta
  \right]. 
  \label{eq:current_I_theta}
\end{split}
\end{equation}
The oscillation frequency at this cone angle is 
\begin{equation}
  f(\theta)
  =
  \frac{\gamma}{2\pi}
  \left[
    H_{\rm appl}
    +
    \left(
      H_{\rm K}
      -
      4\pi M
    \right)
    \cos\theta
  \right].
  \label{eq:frequency}
\end{equation}
Note that the self-oscillation is excited above the critical current $I_{\rm c}=\lim_{\theta \to 0}I(\theta)$, 
\begin{equation}
  I_{\rm c}
  =
  \frac{4 \alpha eMV}{\hbar \eta \lambda}
  \left(
    H_{\rm appl}
    +
    H_{\rm K}
    -
    4\pi M 
  \right), 
  \label{eq:Ic}
\end{equation}
which is $1.6$ mA for $H_{\rm appl}=2.0$ kOe. 
The current dependence of $m_{z}=\cos\theta$ estimated from Eq. (\ref{eq:current_I_theta}) is shown in Fig. \ref{fig:fig3}(c) by the black solid line. 
We also show the maximum and minimum values of $m_{z}$ in the numerical simulations by the red squares and blue circles, respectively. 
It is shown that the analytical theory based on the averaging technique well reproduces 
the exact value of $m_{z}$ estimated from the numerical simulation, particularly in the low-current region. 
We also compare the numerically evaluated oscillation frequency of $m_{x}$ with the analytical theory given by Eq. (\ref{eq:frequency}) 
in Fig. \ref{fig:fig3}(d). 
A good agreement between the numerical simulation (red circles) and the analytical theory (black solid line) is obtained in the low-current region. 
On the other hand, a large difference between them is found in the high-current region, 
except for the good agreement of $m_{z}$ in Fig. \ref{fig:fig3}(c), due to the following reason. 
As mentioned above, the spin torque forces the magnetization to move to the direction antiparallel to $\mathbf{p}=+\mathbf{e}_{x}$. 
With increasing cone angle $\theta$, the projection of the spin torque to the $x$ direction increases, 
which affects the phase of the oscillation around the $z$ axis, as well as the frequency. 
We note that this effect of the spin torque on frequency is not included in Eq. (\ref{eq:frequency}), 
and therefore, the difference between the numerical simulation and analytical theory appears. 


To summarize the above results, the averaging technique works well to study the self-oscillation of magnetization in the low-current region. 
For typical spin-torque oscillators using magnetic tunnel junctions, 
the maximum current available to be applied is about 5 mA, 
which corresponds roughly to 500 mV [\onlinecite{kubota13}]; 
a current larger than this value results in an electrostatic breakdown. 
The available value of the current increases for spin-torque oscillators using a giant magnetoresistive structure. 
It typically becomes 10 mA [\onlinecite{rippard10}]. 
Comparing these values with the results shown in Figs. \ref{fig:fig3}(c) and \ref{fig:fig3}(d), 
we consider that the averaging technique is applicable to the current range available in the experiments. 


We should note that the averaging technique is unnecessary for systems having high symmetry. 
For example, when both the free and pinned layers have perpendicular anisotropy, as in Ref. [\onlinecite{silva10}], 
the LLG equation for $m_{z}$ is independent of the in-plane components, $m_{x}$ and $m_{y}$, 
and the spin and damping torque always cancel each other in the self-oscillation state. 
Then, the solution of $m_{z}$ in the self-oscillation state is exactly constant. 
We also note in Sec. \ref{sec:Generalization of theory} below 
that the averaging technique over a trajectory having a constant cone angle will be generalized 
to that over a constant energy curve. 


\subsection{Rapid relaxation}
\label{sec:Rapid relaxation}

Now let us return to the subject of the relaxation phenomenon in the spin-torque oscillator. 
Usually, the spin-torque asymmetry $\lambda$ is a small parameter, $|\lambda| \ll 1$ [\onlinecite{taniguchi13}]. 
Thus, keeping the lowest order terms of $\lambda$, 
we approximate Eq. (\ref{eq:LLG_averaged}) as 
\begin{equation}
  \frac{d m_{z}}{d \tilde{t}}
  \simeq
  \frac{(1-m_{z}^{2})}{2}
  \left[
    \left(
      2 \alpha
      -
      \lambda 
      h_{\rm s}
    \right)
    m_{z}
    +
    2 \alpha 
    h 
  \right].
  \label{eq:LLG_approx_1}
\end{equation}
The red dashed line in Fig. \ref{fig:fig2} shows Eq. (\ref{eq:LLG_approx_1}), 
indicating that Eq. (\ref{eq:LLG_averaged}) can be well approximated by Eq. (\ref{eq:LLG_approx_1}). 
The stable fixed point estimated from Eq. (\ref{eq:LLG_approx_1}) is given by 
\begin{equation}
  m_{z}^{*}
  =
  \frac{2 \alpha h}{-2 \alpha + \lambda h_{\rm s}}.
  \label{eq:mz_stable}
\end{equation}
Since $|m_{z}^{*}|<1$, the stable fixed point given by Eq. (\ref{eq:mz_stable}) exists when 
$2 \alpha h/(-2 \alpha + \lambda h_{\rm s})<1$, or equivalently, when $I/I_{\rm c}>1$, is satisfied, 
where $I_{\rm c}$ is the critical current to excite the self-oscillation given by Eq. (\ref{eq:Ic}). 
Another condition on the current to excite self-oscillation is summarized in Appendix \ref{sec:AppendixA}. 
For a given $m_{z}^{*}$, the frequency of the self-oscillation is given by Eq. (\ref{eq:frequency}) with $\cos\theta = m_{z}^{*}$. 




\begin{figure}
\centerline{\includegraphics[width=1.0\columnwidth]{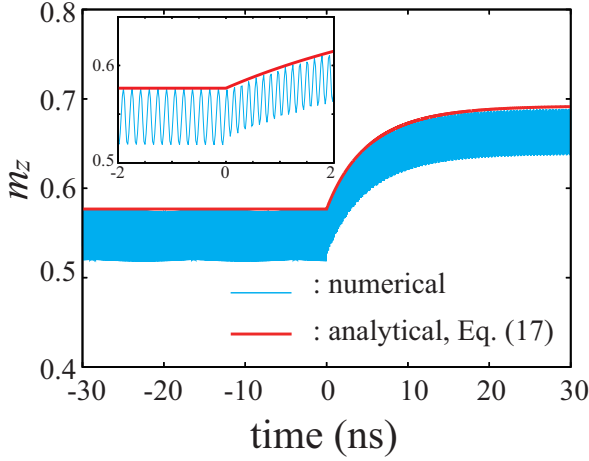}}
\caption{
         The numerical solution of $m_{z}$ obtained from Eq. (\ref{eq:LLG}) is shown by the blue line, 
             where the magnitude of the magnetic field is changed from $H_{0}=2.0$ kOe to $H_{1}=2.4$ kOe at $t=0$. 
             The analytical solution, Eq. (\ref{eq:sol_1}), is also shown by the red line. 
             The inset shows $m_{z}$ near $t=0$. 
         \vspace{-3ex}}
\label{fig:fig4}
\end{figure}




We study the relaxation time using Eq. (\ref{eq:LLG_approx_1}). 
Let us assume that the magnetization is in a certain self-oscillation state when $t \le 0$. 
Denoting the magnetic field for $t \le 0$ as $H_{\rm appl}=H_{0}$, or $h_{0}$ in the dimensionless unit, 
the stable fixed point for $t \le 0$ is given by $m_{z}^{*}(t \le 0) = 2 \alpha h_{0}/(-2 \alpha + \lambda h_{\rm s})$. 
Then, imagine that an additional external field is applied from $t=0$, and the total external magnetic field becomes a different value, 
$H_{\rm appl}=H_{1}$, or $h_{1}$ in the dimensionless unit, 
as shown schematically in Fig. \ref{fig:fig1}(b). 
The magnetization will move to a new stable fixed point 
\begin{equation}
  m_{z}^{*}(t \to \infty)
  =
  \frac{2 \alpha h_{1}}{-2 \alpha + \lambda h_{\rm s}}, 
\end{equation}
where we assume that $H_{1}$ satisfies $2 \alpha h_{1}/(-2 \alpha + \lambda h_{\rm s})<1$. 
When $H_{1}<H_{0}$, $m_{z}^{*}(t \to \infty)$ is smaller than $m_{z}^{*}(t \le 0)$, 
while $m_{z}^{*}(t \to \infty)>m_{z}^{*}(t \le 0)$ when $H_{1}>H_{0}$, as shown in Fig. \ref{fig:fig1}(b). 
This relaxation is described by the following equation obtained from Eq. (\ref{eq:LLG_approx_1}) as 
\begin{equation}
  \frac{d}{d \tilde{t}}
  \delta 
  m_{z}
  \simeq
  -r 
  \delta 
  m_{z}
  -
  u 
  \delta
  m_{z}^{2},
  \label{eq:LLG_approx_2}
\end{equation}
where $\delta m_{z}(t)= m_{z}(t) - m_{z}^{*}(t \to \infty)$, 
whereas $r$ and $u$ are defined as 
\begin{equation}
\begin{split}
  r
  &
  \equiv
  \frac{1}{2}
  \left(
    1
    -
    m_{z}^{* 2}
  \right)
  \left(
    \lambda 
    h_{\rm s}
    -
    2 \alpha
  \right)
\\
  &=
  \frac{1}{2}
  \left[
    1
    -
    \left(
      \frac{2 \alpha h_{1}}{-2 \alpha + \lambda h_{\rm s}}
    \right)^{2}
  \right]
  \left(
    \lambda 
    h_{\rm s}
    -
    2 \alpha
  \right),
  \label{eq:rate_1}
\end{split}
\end{equation}
\begin{equation}
  u
  \equiv
  -m_{z}^{*}
  \left(
    \lambda
    h_{\rm s}
    -
    2 \alpha
  \right)
  =
  -2 \alpha h_{1}.
\end{equation}
Note that $r$ is a positive quantity in the present case because 
$\lambda h_{\rm s}>2 \alpha (1+h_{1})> 2 \alpha$ is satisfied. 
The solution of Eq. (\ref{eq:LLG_approx_2}) is 
\begin{equation}
  m_{z}(t>0)
  =
  m_{z}^{*}(t \to \infty)
  +
  \frac{ \delta m_{z}(0) r e^{-r \tilde{t}}}{r + \delta m_{z}(0) u (1- e^{-r \tilde{t}})},
  \label{eq:sol_1}
\end{equation}
where $\delta m_{z}(0)=m_{z}^{*}(t\le 0) - m_{z}^{*}(t \to \infty)$ is given by 
\begin{equation}
  \delta m_{z}(0)
  =
  \frac{2 \alpha (h_{0}-h_{1})}{-2\alpha + \lambda h_{s}}. 
\end{equation}
Equation (\ref{eq:sol_1}) satisfies $\lim_{t \to 0}m_{z}(t)=m_{z}^{*}(t \le 0)$ 
and $\lim_{t \to \infty}m_{z}(t)=m_{z}^{*}(t \to \infty)$. 
Equation (\ref{eq:sol_1}) indicates that the relaxation occurs exponentially 
within the time scale given by 
\begin{equation}
\begin{split}
  t_{\rm r}
  & \equiv 
  \frac{1}{ \gamma (H_{\rm K}-4\pi M) r}
\\
  &=
  \frac{2}{\gamma (H_{\rm K}-4\pi M) (\lambda h_{\rm s} - 2 \alpha)}
  \left[
    1
    -
    \left(
      \frac{2 \alpha h_{1}}{-2 \alpha + \lambda h_{\rm s}}
    \right)^{2}
  \right]^{-1}.
  \label{eq:relaxation_time}
\end{split}
\end{equation}
The exponential dependence of Eq. (\ref{eq:sol_1}) guarantees a fast relaxation of magnetization. 
Note that the averaging technique of the LLG equation is valid when the oscillation period, Eq. (\ref{eq:frequency}), is shorter than the relaxation time, 
i.e., $1/f \ll t_{\rm r}$. 


One might consider that the exponential dependence of the relaxation is a natural conclusion 
as a result of the existence of the linear term in the LLG equation, Eq. (\ref{eq:LLG_approx_2}). 
However, the linear approximation is no longer applicable near a critical point, 
and the relaxation cannot be described by the exponentials, as shown in Sec. \ref{sec:Critical slowing down} below. 


We confirm the validity of Eq. (\ref{eq:sol_1}) by comparing it with the numerical simulation of Eq. (\ref{eq:LLG}). 
We assume that the magnetic field before $t=0$, $H_{0}$, is 2.0 kOe. 
The field is then changed to $H_{1}=2.4$ kOe at $t=0$. 
The time evolution of $m_{z}(t)$ around $t=0$ obtained by numerically solving Eq. (\ref{eq:LLG}) is 
shown in Fig. \ref{fig:fig4} by the blue line. 
As shown, $m_{z}$ moves to a different state corresponding to the oscillation frequency of 7.6 GHz. 
It can be seen from Fig. \ref{fig:fig4} that Eq. (\ref{eq:sol_1}) well describes the relaxation of magnetization 
from one self-oscillation state to another. 
Both the numerical and analytical solutions indicate that the relaxation occurs within a time on the order of nanoseconds. 
The quantitative value of the relaxation time, Eq. (\ref{eq:relaxation_time}), for the present parameters is 6.3 ns. 
We note that the condition, $1/f \ll t_{\rm r}$, 
to guarantee the validity of Eq. (\ref{eq:relaxation_time}) is quantitatively satisfied. 



\subsection{Current dependences of relaxation time and agility}
\label{sec:Current dependences of relaxation time and agility}



\begin{figure}
\centerline{\includegraphics[width=1.0\columnwidth]{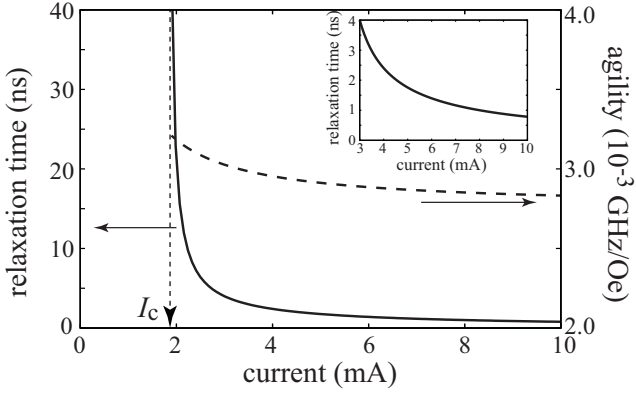}}
\caption{
         The current dependences of the relaxation time and the agility in response to the magnetic field, given by Eqs. (\ref{eq:relaxation_time}) and (\ref{eq:agility_field}), are 
         shown by the solid and dashed lines, respectively. 
         Note that these quantities are defined for the current above $I_{\rm c}\simeq 1.8$ mA for $H_{\rm appl}=2.4$ kOe. 
         The inset shows the relaxation time for $I \gg I_{\rm c}$. 
         \vspace{-3ex}}
\label{fig:fig5}
\end{figure}



The solid line in Fig. \ref{fig:fig5} represents the current dependence of the relaxation time given by Eq. (\ref{eq:relaxation_time}) for $H_{\rm appl}=H_{1}=2.4$ kOe. 
For a large current, $I \gg I_{\rm c} \simeq 1.8$ mA, the relaxation time is on the order of nanoseconds or less, 
guaranteeing a fast relaxation of the magnetization. 
On the other hand, the divergence of the relaxation time near $I_{\rm c}$ indicates the breakdown of the description of relaxation based on exponential dependence. 
This problem is solved in Sec. \ref{sec:Critical slowing down}. 

Another quantity characterizing the relaxation is the agility in response to the magnetic field, 
which is defined as 
\begin{equation}
  \frac{\partial f}{\partial H_{\rm appl}}
  \simeq
  \frac{\gamma}{2\pi}
  \left(
    1
    +
    \frac{2 \alpha}{-2\alpha + \lambda h_{\rm s}}
  \right),
  \label{eq:agility_field}
\end{equation}
where we use Eqs. (\ref{eq:frequency}) and (\ref{eq:mz_stable}) 
(see also Appendix \ref{sec:AppendixB}). 
Equation (\ref{eq:agility_field}) is independent of the magnetic field. 
We also note that the agility is practically independent of the current, as can be seen in Fig. \ref{fig:fig5}. 
These results indicate that the frequency shift by the relaxation is solely determined by the difference of the magnetic field, $\Delta H=|H_{0}-H_{1}|$. 



\subsection{Critical slowing down}
\label{sec:Critical slowing down}

\begin{figure}
\centerline{\includegraphics[width=1.0\columnwidth]{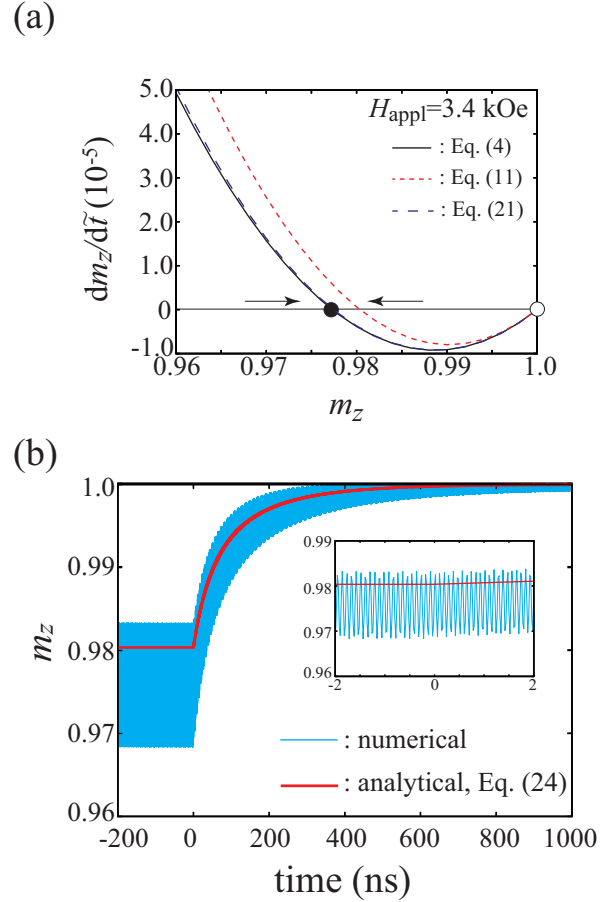}}
\caption{
         (a) The time derivative obtained from Eqs. (\ref{eq:LLG_averaged}), (\ref{eq:LLG_approx_1}), and (\ref{eq:LLG_approx_3}) (blue dashed) respectively,
             for the case with $I=2.5$ mA and $H_{\rm appl}=3.4$ kOe. 
         (b) The blue line is the numerical solution of $m_{z}$ obtained from Eq. (\ref{eq:sol_2}),  
             where $H_{0}=3.4$ kOe and $H_{1}=3.5$ kOe at $t=0$. 
             The red line is the analytical solution Eq. (\ref{eq:sol_2}). 
             The inset shows $m_{z}$ near $t=0$. 
         \vspace{-3ex}}
\label{fig:fig6}
\end{figure}

The fixed point is called a critical point when a stable fixed point comes close to an unstable one. 
This case happens when a condition, $I/I_{\rm c} \simeq 1$, is satisfied. 
In this case, Eq. (\ref{eq:LLG_averaged}) is well approximated by expanding it around $m_{z}=1$ as 
\begin{equation}
  \frac{d}{d \tilde{t}}
  m_{z}
  \simeq
  -r^{\prime}
  \left(
    m_{z}
    -
    1
  \right)
  -
  u^{\prime}
  \left(
    m_{z}
    -
    1
  \right)^{2},
  \label{eq:LLG_approx_3}
\end{equation}
where $r^{\prime}$ and $u^{\prime}$ are given by 
\begin{equation}
  r^{\prime}
  =
  2 \alpha
  \left(
    1
    +
    h
  \right)
  -
  \lambda
  h_{\rm s},
  \label{eq:rate_2}
\end{equation}
\begin{equation}
  u^{\prime}
  =
  \alpha
  \left(
    3
    +
    h
  \right)
  -
  \frac{3 \lambda (1-\lambda^{2})}{2}
  h_{\rm s}.
  \label{eq:rate_u_prime}
\end{equation}
Figure \ref{fig:fig6}(a) shows Eq. (\ref{eq:LLG_averaged}), (\ref{eq:LLG_approx_1}), and (\ref{eq:LLG_approx_3}) 
by the black solid, red dotted, and blue dashed lines, respectively, 
for the case of $I=2.5$ mA and $H_{\rm appl}=3.4$ kOe. 
In this case, $I/I_{\rm c} \simeq 1$, and the stable fixed point becomes close to the unstable one. 
As shown, Eq. (\ref{eq:LLG_approx_3}) well reproduces the exact equation (\ref{eq:LLG_averaged}), than Eq. (\ref{eq:LLG_approx_1}), 
indicating that Eq. (\ref{eq:LLG_approx_3}) is useful for investigating the relaxation time near the critical point. 
The fixed points obtained from Eq. (\ref{eq:LLG_approx_3}) are $m_{z}=1$ and $m_{z}^{*}=1-(r^{\prime}/u^{\prime})$, 
where the former corresponds to the unstable fixed point, 
and the latter is the stable fixed point when $1-(r^{\prime}/u^{\prime})<1$ is satisfied.



Let us investigate the relaxation time near the critical point. 
The solution of Eq. (\ref{eq:LLG_approx_3}) is given by 
\begin{equation}
  m_{z}(t>0)
  =
  1
  +
  \frac{\delta m_{z}^{\prime}(0) r^{\prime} e^{-r^{\prime} \tilde{t}}}{r^{\prime} + \delta m_{z}^{\prime}(0) u^{\prime} ( 1 - e^{-r^{\prime} \tilde{t}})},
  \label{eq:sol_2}
\end{equation}
where $\delta m_{z}^{\prime}(0)=m_{z}^{*}(t \le 0)-1$. 
The value of the magnetic field in $r^{\prime}$ and $u^{\prime}$ in Eq (\ref{eq:sol_2}) should be regarded as that for $t>0$. 
We note that 
\begin{equation}
  \lim_{t \to \infty}
  m_{z}(t)
  =
  \begin{cases}
    1 & (r^{\prime}>0) \\
    1-(r^{\prime}/u^{\prime}) & (r^{\prime}<0)
  \end{cases},
  \label{eq:case}
\end{equation}
where the upper case ($r^{\prime}>0$) corresponds to the condition necessary to excite self-oscillation, $I/I_{\rm c}>1$, is no longer satisfied,
and therefore, magnetization moves to an energetically stable state. 
The lower case ($r^{\prime}<0$) in Eq. (\ref{eq:case}) corresponds to the condition where $I/I_{\rm c}>1$ is satisfied, 
and magnetization moves to a stable fixed point $m_{z}^{*}=1-(r^{\prime}/u^{\prime})$. 
In particular, near the critical point where $r^{\prime} \simeq 0$, or equivalently $I/I_{\rm c} \simeq 1$, 
Eq. (\ref{eq:sol_2}) is approximated to 
\begin{equation}
  \lim_{r^{\prime} \to 0}
  m_{z}(t>0)
  =
  1
  +
  \frac{\delta m_{z}^{\prime}(0)}{1 + \delta m_{z}^{\prime}(0) u^{\prime} \tilde{t}}, 
  \label{eq:sol_3}
\end{equation}
with $u^{\prime}\to 3 \alpha \lambda^{2}(1+h_{1})-2\alpha h_{1}$, 
which is negative when the condition on the magnetic field necessary to excite self-oscillation is satisfied 
[see Eq. (\ref{eq:field_condition}) in Appendix \ref{sec:AppendixA}].
Equation (\ref{eq:sol_3}) shows that the magnetization relaxation near the critical point is inversely proportional to the time, 
which is much slower than the exponential relaxation seen far from the critical point [Eq. (\ref{eq:sol_1})]. 
The phenomenon is similar to the critical slowing down found near phase transitions, 
where the relaxation is described by algebraic functions, rather than exponentials [\onlinecite{strogatz01}]. 


The critical slowing down is confirmed from the numerical simulation of Eq. (\ref{eq:LLG}). 
Figure \ref{fig:fig6}(b) shows the time evolution of $m_{z}$ obtained from the numerical simulation of Eq. (\ref{eq:LLG}) by the blue line, 
where the applied magnetic field is changed from $H_{0}=3.4$ kOe to $H_{1}=3.5$ kOe at $t=0$. 
Magnetization for these magnetic fields occurs near the critical point, 
because the value of the field satisfying $I=I_{\rm c}$ with $I=2.5$ mA is $H_{\rm appl}=3.45$ kOe. 
The analytical solution, Eq. (\ref{eq:sol_2}), is also shown by the red line. 
Both the numerical and analytical results indicate that the relaxation occurs over a time period longer than 100 ns, 
which is much slower than that shown in Fig. \ref{fig:fig4}. 
The good agreement between the numerical and analytical results in Fig. \ref{fig:fig6}(b) also indicates that 
the critical slowing down occurs in the spin-torque oscillator. 


A large current is necessary to excite a self-oscillation at a stable fixed point far away from the unstable one, 
whereas the current stabilizing the oscillation near the critical point is small. 
The excitation of self-oscillation near the critical point is therefore preferable to reduce both the current magnitude and power consumption. 
The above results, however, suggest that use of self-oscillation near the critical point should be avoided 
for rapid operation of the spin-torque oscillator. 




\section{Generalization of theory}
\label{sec:Generalization of theory}

The theory developed above focuses on a spin-torque oscillator consisting of a perpendicularly magnetized free layer 
and an in-plane magnetized pinned layer. 
In this section, we generalize the description of the relaxation, 
and show that the critical slowing down appears in general cases. 


\subsection{LLG equation}
\label{sec:LLG equation}

We start from the LLG equation of Eq. (\ref{eq:LLG}), 
Now, however, the magnetic field $\mathbf{H}$, 
the spin-torque strength $H_{\rm s}$, and the pinned layer magnetization $\mathbf{p}$ are assumed to be arbitrary. 
A self-oscillation is excited when the energy supplied by the spin torque balances the dissipation due to the damping, 
and therefore, the magnetic energy is almost constant during a precession. 
Let us denote the energy density as $E$, which is related to the magnetic field via $E=-M \int d \mathbf{m}\cdot\mathbf{H}$ [\onlinecite{lifshitz80}]. 
Using Eq. (\ref{eq:LLG}), the energy change is described by 
$dE/dt=-M \mathbf{H}\cdot(d \mathbf{m}/dt)$. 
Since the energy $E$ changes slowly in the self-oscillation state, 
it is a good approximation to average the equation $dE/dt$ over a constant energy curve of $E$. 
We then obtain 
\begin{equation}
  \frac{1}{\tau(E)}
  \oint
  dt 
  \frac{dE}{dt}
  =
  \frac{1}{\tau(E)}
  \left[
    \mathscr{W}_{\rm s}(E)
    +
    \mathscr{W}_{\alpha}(E)
  \right],
  \label{eq:LLG_energy}
\end{equation}
where the integral is over a precession period of a constant energy curve of $E$. 
The work done by the spin torque and dissipation due to the damping torque during a precession are denoted as 
$\mathscr{W}_{\rm s}(E)$ and $\mathscr{W}_{\alpha}(E)$, respectively, 
and are defined as 
\begin{equation}
  \mathscr{W}_{\rm s}(E)
  =
  \oint
  dt 
  \gamma
  M H_{\rm s}
  \left[
    \mathbf{p}
    \cdot
    \mathbf{H}
    -
    \left(
      \mathbf{m}
      \cdot
      \mathbf{p}
    \right)
    \left(
      \mathbf{m}
      \cdot
      \mathbf{H}
    \right)
  \right],
  \label{eq:W_s}
\end{equation}
\begin{equation}
  \mathscr{W}_{\alpha}(E)
  =
  -\oint 
  dt 
  \alpha 
  \gamma 
  M 
  \left[
    \mathbf{H}^{2}
    -
    \left(
      \mathbf{m}
      \cdot
      \mathbf{H}
    \right)^{2}
  \right],
  \label{eq:W_alpha}
\end{equation}
where we neglect higher-order terms of $\alpha$ and $H_{\rm s}$. 
The precession period is 
\begin{equation}
  \tau
  =
  \oint dt, 
\end{equation}
which relates to the frequency of the self-oscillation via $f=1/\tau$. 
Note that $\mathscr{W}_{\rm s}$, $\mathscr{W}_{\alpha}$, and $\tau$ are functions of 
energy density $E$ corresponding to a self-oscillation state. 
In other words, the self-oscillation state is identified by $E$. 
In the following, we denote the left hand side of Eq. (\ref{eq:LLG_energy}) as $dE/dt$, for simplicity. 
The condition for sustaining self-oscillation is now generalized to $d E/dt=0$. 
We note that the averaging technique of the LLG equation on a constant energy curve might show a different behavior with the numerical result 
for particular limits, as discussed in Sec. \ref{sec:Validity of the averaging technique}. 


The averaging technique of the slow variable was introduced in previous research [\onlinecite{bertotti09,apalkov05}] 
to study the magnetization dynamics in nanostructured ferromagnets. 
The steady state solution of Eq. (\ref{eq:LLG_energy}) was shown to be useful for building the phase portrait of the magnetization dynamics 
at zero temperature [\onlinecite{bertotti09}]. 
It was also shown that the Fokker-Planck equation based on Eq. (\ref{eq:LLG_energy}) can be used to evaluate 
the thermally activated magnetization reversal rate at finite temperature [\onlinecite{apalkov05}]. 
On the other hand, in this study, Eq. (\ref{eq:LLG_energy}) is used to investigate the relaxation phenomenon in the spin-torque oscillator. 
Equation (\ref{eq:LLG_energy}) describes the energy change by the competition between the spin and damping torques 
when its time scale is slower than the oscillation period. 



In the previous section, we used $m_{z}$, instead of $E$, 
because these are directly related as $E=-MH_{\rm appl}m_{z}-[M (H_{\rm K}-4\pi M)/2]m_{z}^{2}$. 
The explicit forms of $\mathscr{W}_{\rm s}$ and $\mathscr{W}_{\alpha}$ for the system considered in the previous section are 
\begin{equation}
\begin{split}
  \frac{\mathscr{W}_{\rm s}}{\tau}
  =&
  \frac{\gamma \hbar \eta I}{2e \lambda V}
  \left[
    \frac{1}{\sqrt{1-\lambda^{2}(1-m_{z}^{2})}}
    -
    1
  \right]
\\
  & 
  \times
  \left[
    H_{\rm appl}
    +
    \left(
      H_{\rm K}
      -
      4\pi M 
    \right)
    m_{z}
  \right]
  m_{z},
  \label{eq:W_s_example}
\end{split}
\end{equation}
\begin{equation}
\begin{split}
  \frac{\mathscr{W}_{\alpha}}{\tau}
  =
  -\alpha 
  \gamma
  M 
  \left[
    H_{\rm appl}
    +
    \left(
      H_{\rm K}
      -
      4\pi M
    \right)
    m_{z}
  \right]^{2}
  \left(
    1
    -
    m_{z}^{2}
  \right),
  \label{eq:W_alpha_example}
\end{split}
\end{equation}
where $\tau=2\pi/\{\gamma [H_{\rm appl}+(H_{\rm K}-4\pi M)m_{z}]\}$. 
Using the relation $dE/dt=(dE/dm_{z})(dm_{z}/dt)$ 
and Eqs. (\ref{eq:W_s_example}) and (\ref{eq:W_alpha_example}), 
it can be confirmed that Eq. (\ref{eq:LLG_energy}) reproduces Eq. (\ref{eq:LLG_averaged}). 


In general, the values of $\mathscr{W}_{\rm s}$ and $\mathscr{W}_{\alpha}$ depend on the magnetic anisotropy, 
the magnitude and direction of the external field, the relative angle of the magnetizations, and so on. 
The analytical formulas of $\mathscr{W}_{\rm s}$ and $\mathscr{W}_{\alpha}$ have been derived exactly or approximately in several cases. 
[\onlinecite{serpico03,bertotti04,bertotti05,taniguchi13PRB,newhall13,lee14,pinna14,taniguchi15PRB,taniguchi15,taniguchi16PRB,taniguchi16}]. 
However, unless the system has some symmetry such as an axial symmetry of the magnetic anisotropy, 
it is usually difficult to derive the analytical formulas. 
The numerical evaluations of the integrals in Eqs. (\ref{eq:W_s}) and (\ref{eq:W_alpha}) are useful in such cases, 
where the constant energy curve is calculated by solving the Landau-Lifshitz equation, $d \mathbf{m}/dt=-\gamma \mathbf{m} \times \mathbf{H}$. 


The critical current $I_{\rm c}$ to excite a self-oscillation is defined as a current 
satisfying $\mathscr{W}_{\rm s}(E_{\rm min})+\mathscr{W}_{\alpha}(E_{\rm min})=0$, 
where $E_{\rm min}$ is the energy density corresponding to the minimum energy state. 
We note that both $\mathscr{W}_{\rm s}(E_{\rm min})$ and $\mathscr{W}_{\alpha}(E_{\rm min})$ are zero 
when the minimum energy state is a point, as in the case of the perpendicular ferromagnet described in the previous section. 
This is because the constant energy curve is just a point, and therefore, even though the period $\tau$ is finite, 
the integrals of Eqs. (\ref{eq:W_s}) and (\ref{eq:W_alpha}) become zero. 
This can also be confirmed from Eqs. (\ref{eq:W_s_example}) and (\ref{eq:W_alpha_example}), where the minimum energy state corresponds to $|m_{z}|=1$. 
When both $\mathscr{W}_{\rm s}$ and $\mathscr{W}_{\alpha}$ are zero at $E=E_{\rm min}$, 
the minimum energy state is always a fixed point. 
However, the ratio $\lim_{E \to E_{\rm min}}\mathscr{W}_{\rm s}(E)/\mathscr{W}_{\alpha}(E)$ is finite, and the critical current $I_{\rm c}$ is well-defined. 
The critical current $I_{\rm c}$ is given by 
\begin{equation}
  \lim_{E \to E_{\rm min}}
  \frac{\mathscr{W}_{\rm s}(E)}{\mathscr{W}_{\alpha}(E)}
  =
  \frac{d \mathscr{W}_{\rm s}/dE}{d \mathscr{W}_{\alpha}/dE}
  \bigg|_{E=E_{\rm min}}
  =
  -\frac{I}{I_{\rm c}},
  \label{eq:Ic_def_general}
\end{equation}
where we use $H_{\rm s} \propto I$. 
The minimum energy state is stable when $I/I_{\rm c}<1$. 
On the other hand, self-oscillation is excited when $I/I_{\rm c}>1$. 
In this case, a stable fixed point appears at a higher energy state, 
whereas the minimum energy state becomes the unstable one. 
Therefore, the minimum energy state in spin-torque oscillators can be classified into a transcritical bifurcation [\onlinecite{strogatz01}]. 




\begin{figure}
\centerline{\includegraphics[width=1.0\columnwidth]{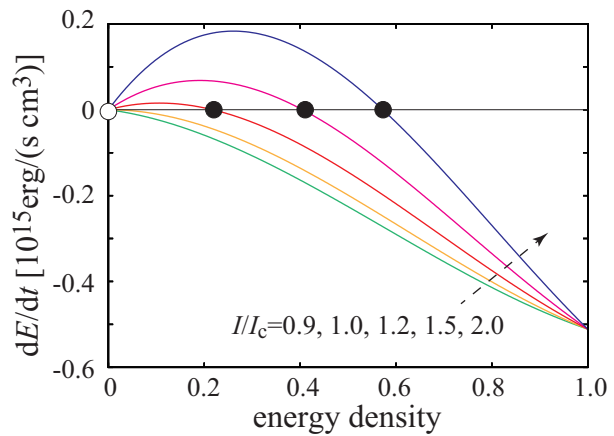}}
\caption{
         Dependences of $dE/dt=(\mathscr{W}_{\rm s}+\mathscr{W}_{\alpha})/\tau$ on the several values of current $I$. 
         The energy density $E$ in the horizontal axis is normalized as $(E-E_{\rm min})/(E_{K}-E_{\rm min})$, where $E_{\rm K}=E(m_{z}=0)$. 
         The black circles correspond to the stable fixed points, whereas the white circle is the unstable one. 
         The magnetic field is $H_{\rm appl}=2.0$ kOe, corresponding to $I_{\rm c}\simeq 1.6$ mA. 
         \vspace{-3ex}}
\label{fig:fig7}
\end{figure}



The discussion in this section is applicable to any type of spin-torque oscillator when the above assumptions hold. 
It is, however, useful to show an example of $dE/dt=(\mathscr{W}_{\rm s}+\mathscr{W}_{\alpha})/\tau$ 
to get hold of the overall picture. 
Therefore, we show $dE/dt$ with Eqs. (\ref{eq:W_s_example}) and (\ref{eq:W_alpha_example}) in Fig. \ref{fig:fig7}, 
where $H_{\rm appl}=2.0$ kOe, corresponding to $I_{\rm c} \simeq 1.6$ mA. 
The figure shows that $dE/dt$ has two fixed points, indicated by the black and white circles, when $I/I_{\rm c}>1$. 
The black circle corresponds to the stable fixed point and the white to the unstable. 


Let us assume that the spin-torque oscillator shows a self-oscillation on a constant energy curve of $E=E_{0}$ for $t<0$, 
and that from $t>0$, it relaxes to a stable fixed point at $E=E_{1}$ 
due to the change of the current and/or field. 
The necessary condition to stabilize the self-oscillation for $t>0$ is $\mathscr{W}_{\rm s}(E_{1})+\mathscr{W}_{\alpha}(E_{1})=0$. 
Then, the time evolution of $\delta E=E-E_{1}$ is described by 
\begin{equation}
  \frac{d}{dt}
  \delta 
  E 
  \simeq
  -\mathscr{R}
  \delta E
  -
  \mathscr{U}
  \delta E^{2},
  \label{eq:LLG_energy_approx}
\end{equation}
where $\mathscr{R}$ and $\mathscr{U}$ are defined as 
\begin{equation}
  \mathscr{R}
  =
  -\frac{1}{\tau}
  \frac{d}{dE}
  \left(
    \mathscr{W}_{\rm s}
    +
    \mathscr{W}_{\alpha}
  \right)
  \bigg|_{E=E_{1}},
  \label{eq:r_general}
\end{equation}
\begin{equation}
  \mathscr{U}
  =
  -\frac{1}{2}
  \frac{d^{2}}{dE^{2}}
  \frac{1}{\tau}
  \left(
    \mathscr{W}_{\rm s}
    +
    \mathscr{W}_{\alpha}
  \right)
  \bigg|_{E=E_{1}}.
  \label{eq:u_general}
\end{equation}
The quantity $\mathscr{R}$ at the stable fixed point is positive 
when the condition to excite a self-oscillation, $I/I_{\rm c}>1$, is satisfied, 
and therefore, the constant energy curve of $E_{1}(>E_{\rm min})$ is an attractor. 
The positive sign of $\mathscr{R}$ can also be understood from Fig. \ref{fig:fig7}, 
where $\mathscr{R}$ corresponds to the gradient of $dE/dt$ at the black circles multiplied by a negative sign. 
The solution of Eq. (\ref{eq:LLG_energy_approx}) in this case is 
\begin{equation}
  \delta 
  E(t)
  =
  \frac{\delta E(0) \mathscr{R} e^{-\mathscr{R}t}}{\mathscr{R} + \delta E(0) \mathscr{U}(1-e^{-\mathscr{R}t})}, 
  \label{eq:E_solution_1}
\end{equation}
where $\delta E(0)=E_{0}-E_{1}$. 
Thus, the magnetization relaxes exponentially versus the time to the self-oscillation state within the time scale $1/\mathscr{R}$. 


\subsection{Critical slowing down in general cases}
\label{sec:Critical slowing down in general cases}

In the section above, we investigated the time evolution of $\delta E$ near the stable fixed point $E_{1}$. 
Here, let us consider the case near the minimum energy state. 
As already mentioned, the minimum energy state is the stable (unstable) fixed point when $I/I_{\rm c}<(>)1$. 
We note that $\delta E$ near the minimum energy state obeys an equation similar to Eq. (\ref{eq:LLG_energy_approx}), 
but that $\mathscr{R}$ and $\mathscr{U}$ are replaced by 
\begin{equation}
\begin{split}
  \mathscr{R}^{\prime}
  &=
  -\frac{1}{\tau}
  \frac{d}{dE}
  \left(
    \mathscr{W}_{\rm s}
    +
    \mathscr{W}_{\alpha}
  \right)
  \bigg|_{E=E_{\rm min}}
\\
  &=
  \frac{1}{\tau}
  \left(
    -\frac{d \mathscr{W}_{\alpha}}{dE}
  \right)
  \bigg|_{E=E_{\rm min}}
  \left(
    1
    -
    \frac{I}{I_{\rm c}}
  \right),
  \label{eq:r_unstable}
\end{split}
\end{equation}
\begin{equation}
  \mathscr{U}^{\prime}
  =
  -\frac{1}{2}
  \frac{d^{2}}{dE^{2}}
  \frac{1}{\tau}
  \left(
    \mathscr{W}_{\rm s}
    +
    \mathscr{W}_{\alpha}
  \right)
  \bigg|_{E=E_{\rm min}},
\end{equation}
respectively, 
where we use Eq. (\ref{eq:Ic_def_general}). 
We note that $\mathscr{R}^{\prime}$ is positive (negative) when $I/I_{\rm c}<(>)1$. 
This can be understood from the fact that $-d \mathscr{W}_{\alpha}/dE$ is positive near the minimum energy state, 
according to the definition of $\mathscr{W}_{\alpha}$, 
i.e., $\mathscr{W}_{\alpha}=0$ at the minimum energy state 
and $\mathscr{W}_{\alpha}<0$ for higher energy states because $\mathscr{W}_{\alpha}$ is the dissipation. 
The sign of $\mathscr{R}^{\prime}$ can also be understood from Fig. \ref{fig:fig7}, 
where $\mathscr{R}^{\prime}$ corresponds to the gradient of $dE/dt$ at the minimum energy state (white circle) multiplied by a negative sign. 
When $I/I_{\rm c}>1$, the gradient is positive, i.e., $\mathscr{R}^{\prime}<0$, 
whereas the gradient is negative for $I/I_{\rm c}<1$, corresponding to $\mathscr{R}^{\prime}>0$. 


The solution of $\delta E$ near the minimum energy state is also given by Eq. (\ref{eq:E_solution_1}) with $\mathscr{R}^{\prime}$ and $\mathscr{U}^{\prime}$. 
The fact that $\mathscr{R}^{\prime}<0$ for $I/I_{\rm c}>1$ means that 
the minimum energy state is unstable due to the energy supplied by the spin torque, 
and that $\delta E$ moves exponentially versus time to the stable fixed point showing self-oscillation. 


On the other hand, when $I/I_{\rm c} \simeq 1$, the stable and unstable fixed points approach each other. 
In this case, $\mathscr{R} \to \mathscr{R}^{\prime}$, 
but since $\mathscr{R}$ and $\mathscr{R}^{\prime}$ have different energy dependences, 
$\mathscr{R} \to 0$. 
It can be seen from Eq. (\ref{eq:r_unstable}) that $\mathscr{R}^{\prime} \simeq 0$ in this case, 
and, the solution of $\delta E$ is given by 
\begin{equation}
  \delta
  E
  =
  \frac{\delta E(0)}{1+\delta E(0) \mathscr{U}^{\prime} t}.
  \label{eq:E_solution_2}
\end{equation}
Equation (\ref{eq:E_solution_2}) indicates that the relaxation of $\delta E$ near the critical point is slow. 
This is the derivation of the critical slowing down for arbitrary types of spin-torque oscillators. 


In the derivation of Eq. (\ref{eq:E_solution_2}), we assume that $\mathscr{U}^{\prime} \neq 0$ at the critical point. 
This assumption is identical to that of $d^{2} (\mathscr{W}_{\rm s}+\mathscr{W}_{\alpha})/ E^{2} \neq 0$, 
where we use $\mathscr{W}_{\rm s}+\mathscr{W}_{\alpha}=0$ and $\mathscr{R} \propto d (\mathscr{W}_{\rm s}+\mathscr{W}_{\alpha})/dE=0$ at the critical point. 
One might consider a general case in which both $\mathscr{R}$ and $\mathscr{U}$ are zero at the critical point. 
Let us assume that $\delta E$ obeys the following equation 
\begin{equation}
  \frac{d}{dt}
  \delta E 
  =
  -\mathscr{G}_{n}
  \delta E^{n}, 
  \label{eq:LLG_energy_general}
\end{equation}
where $n$ is a positive integer ($n \in \mathbb{N}$), 
while $\mathscr{G}_{n}$ is the $n$-th order expansion coefficient of Eq. (\ref{eq:LLG_energy}), 
\begin{equation}
  \mathscr{G}_{n}
  =
  -\frac{1}{n!}
  \frac{d^{n}}{dE^{n}}
  \frac{1}{\tau}
  \left(
    \mathscr{W}_{\rm s}
    +
    \mathscr{W}_{\alpha}
  \right). 
\end{equation}
Here, we assume that $\mathscr{G}_{\ell}=0$ for $\ell <n$. 
The solution of $\delta E$ is 
\begin{equation}
  \delta 
  E
  =
  \left[
    \delta 
    E^{1-n}(0)
    +
    \left(
      n
      -
      1
    \right)
    \mathscr{G}_{n}
    t
  \right]^{-1/(n-1)}.
  \label{eq:E_solution_3}
\end{equation}
The solution of $\delta E$ becomes the exponential $\delta E=\delta E(0) e^{-\mathscr{G}_{1}t}$ when $n=1$ 
corresponding to the case $\mathscr{R} \neq 0$ in Eq. (\ref{eq:LLG_energy_approx}). 
On the other hand, when $\mathscr{R} = 0$ ($\mathscr{G}_{1}=0$), 
the solution of $\delta E$ behaves as 
\begin{equation}
  \delta E 
  \sim 
  t^{-1/(n-1)}. 
\end{equation}
This solution indicates that the relaxation near the critical point is described by algebraic functions, rather than exponentials, for general cases. 
Therefore, the critical slowing down appears even if $\mathscr{U}^{\prime}=0$ at the critical point. 
For the critical slowing down, $\delta E(0)>0$ because energy $E$ moves to the minimum energy state. 
Then, the least nonzero coefficient $\mathscr{G}_{n}$ ($n \ge 2$) near the critical point is positive to guarantee a monotonic relaxation of $\delta E$ [\onlinecite{comment1}]. 


\subsection{Discussion}
\label{sec:Discussion}

Equations (\ref{eq:E_solution_1}), (\ref{eq:E_solution_2}), and (\ref{eq:E_solution_3}) provide general descriptions 
of the relaxation phenomena in spin-torque oscillators. 
Equation (\ref{eq:E_solution_1}) guarantees a fast relaxation to the self-oscillation state obeying the exponential law. 
Equation (\ref{eq:r_general}) implies that a large current results in a fast relaxation 
because, roughly speaking, the relaxation rate $\mathscr{R}$ is proportional to the current $I$ through $\mathscr{W}_{\rm s} \propto I$. 
On the other hand, Eqs. (\ref{eq:E_solution_2}) and (\ref{eq:E_solution_3}) indicate the existence of the critical slowing down in general cases, 
where the relaxation is slow. 
The critical slowing down has not been investigated in the spin-torque oscillators 
but is often found in the phase transition, where the relaxation time to an equilibrium becomes infinite. 
In fact, the equation of motion, for example Eq. (\ref{eq:LLG_energy_approx}), can be found in the other nonlinear systems 
such as the laser threshold and chemical reactions [\onlinecite{strogatz01}]. 


The relaxation near the phase relaxation is characterized by the dynamical critical exponent $z$, 
which is defined as $\mathscr{R}^{\prime} \propto (I_{\rm c}-I)^{z}$. 
Equation (\ref{eq:r_unstable}) indicates that the dynamical critical exponent $z$ is one. 
In fact, Eq. (\ref{eq:rate_2}) can be expressed as $r^{\prime}=2 \alpha (1+h)(1 - I/I_{\rm c})^{z=1}$. 
We note that $z=1$ is valid near the critical point only. 
On the other hand, the current dependence of the rate $\mathscr{R}$ at the stable fixed point cannot be described by such a simple form. 
For example, Eq. (\ref{eq:rate_1}) is proportional to $1-[(-2\alpha + \lambda h_{\rm c})/(-2\alpha + \lambda h_{\rm s})]^{2}$, 
with $h_{\rm c}/h_{\rm s}=I_{\rm c}/I$, which is clearly different from the current dependence of $(I/I_{\rm c}-1)^{z=1}$. 
We also note that a different type of critical exponent for spin torque was discussed in a previous study 
of the thermally activated magnetization switching [\onlinecite{taniguchi13PRB}]. 
The differences between the present and previous works are summarized in Appendix \ref{sec:AppendixC}. 


The above theory is valid when the period of self-oscillation $\tau$ is shorter than the relaxation time, $1/\mathscr{R}$. 
This means that $- d (\mathscr{W}_{\rm s} + \mathscr{W}_{\alpha})/dE \ll 1$ at the fixed points. 
As mentioned, this condition is satisfied in the previous section, 
where the values of the parameters are derived from the experiment [\onlinecite{kubota13}]. 
The recent experiments on the relaxation of the in-plane magnetized free layer under magnetic pulses 
where the oscillation frequency, $1/\tau$, is about 3 GHz whereas the relaxation occurs in nanoseconds, 
also implies that this condition is satisfied [\onlinecite{suto11,nagasawa11}]. 
Therefore, the theory developed here can presumably be applied widely in this field. 
An exception is the perpendicularly magnetized spin-Hall system without a magnetic field, 
where a high symmetry results in $\mathscr{W}_{\rm s}=0$, and thus the averaging technique of the LLG equation is no longer applicable [\onlinecite{taniguchi15PRB}]. 


The critical slowing down found in this paper is related not only to the self-oscillation but also to magnetization switching. 
The analytical theory of the switching time was developed in Ref. [\onlinecite{sun00}] by focusing on 
the instability near the unstable fixed point. 
This work uses a linear approximation, 
corresponding to the limits of $E \to E_{\rm min}$, $\mathscr{G}_{1} \neq 0$, and neglecting the higher order terms of $\delta E$ in Eq. (\ref{eq:LLG_energy_general}). 
Thus, critical slowing down was not found. 
An exactly solvable problem [\onlinecite{yamada15}] on switching time, as well as the approximated solution at finite temperature [\onlinecite{tomita11}], have also been reported, 
where again the critical slowing down was not found. 
(details regarding the definition of the switching time in Ref. [\onlinecite{yamada15}] are summarized in Appendix \ref{sec:AppendixD}). 
It is preferable to switch magnetization by low current to reduce power consumption. 
It has been shown, however, that the switching time becomes longer when the current applied to the free layer is close to the critical current [\onlinecite{taniguchi15PRB}]. 
This result can be explained in terms of the critical slowing down, where $\mathscr{R}^{\prime} \sim 0$ for $I/I_{\rm c} \simeq 1$, 
and therefore, the relaxation becomes slow. 



\section{Conclusion}
\label{sec:Conclusion}

In conclusion, we studied relaxation time to the self-oscillation state in a spin-torque oscillator theoretically. 
The analytical formula for relaxation time, characterizing the exponential relaxation to the self-oscillation state, 
was derived by solving the LLG equation. 
The validity of the derived formula was confirmed by comparison with a numerical simulation. 
Both the analytical and numerical calculations showed that the relaxation time is on the order of nanoseconds 
when the oscillator is far away from the critical point. 
On the other hand, a critical slowing down appeared near the critical point, 
where the relaxation was inversely proportional to the time 
and was on the order of hundreds of nanoseconds. 
Here, it was shown that the linear approximation to the LLG equation is no longer applicable, 
and a nonlinear analysis based on the theory of phase transition is necessary to clarify the relaxation phenomena. 
The theoretical formulas were derived for a spin-torque oscillator consisting of 
a perpendicularly magnetized free layer and an in-plane magnetized pinned layer, 
and then were further developed so that they could be applied in arbitrary types of spin-torque oscillators. 
The dynamical critical exponent of the phase transition between the critical point and the self-oscillation state was found to be one. 
These results provide a comprehensive description of 
the relaxation and critical phenomena in spin-torque oscillators. 


\section*{Acknowledgement}

The authors are grateful to Takehiko Yorozu and Tazumi Nagasawa for valuable discussions. 
T. T. is thankful to Satoshi Iba, Aurelie Spiesser, Hiroki Maehara, and Ai Emura for their support and encouragement. 
T. T. is supported by JSPS KAKENHI Grant-in-Aid for Young Scientists (B) 16K17486. 
Y. U. is supported by JSPS KAKENHI Grants. No. JP26220711 and No. 26400390. 



\appendix

\section{The upper limit of the current to stabilize self-oscillation}
\label{sec:AppendixA}

Equation (\ref{eq:Ic}) represents the critical current to destabilize the magnetization in equilibrium. 
Self-oscillation is excited when the condition $I/I_{\rm c}>1$ is satisfied. 
In this sense, $I_{\rm c}$ determines the lower limit of the current to excite self-oscillation. 
On the other hand, when current $I$ becomes sufficiently large, 
the large spin torque forces the magnetization direction to be fixed in the film plane, 
and the self-oscillation is no longer excited. 
This fact indicates that there is an upper limit of the current to excite self-oscillation. 


The upper limit of the current to excite the self-oscillation can be approximately determined as follows. 
When the current is large, the magnetization direction is fixed to the $xy$ plane, 
and thus, $\mathbf{m}$ can be expressed as $\mathbf{m}=(\cos\phi,\sin\phi,0)$. 
Also, $d \mathbf{m}/dt$ in Eq. (\ref{eq:LLG}) is zero when $\mathbf{m}$ is fixed. 
Then, we find that 
\begin{equation}
  H_{\rm appl}
  +
  H_{\rm s}
  \sin
  \phi
  =
  0.
\end{equation}
The solution of $\phi$ is 
\begin{equation}
  \phi
  \simeq 
  \sin^{-1}
  \left[
    \frac{H_{\rm appl}}{H_{\rm s}^{(0)}}
  \right],
\end{equation}
where we neglect the small parameter $\lambda$ and introduce 
\begin{equation}
  H_{\rm s}^{(0)}
  =
  \frac{\hbar \eta I}{2eMV}. 
\end{equation}
The solution of $\phi$ exists when $|H_{\rm appl}/H_{\rm s}^{(0)}| \le 1$. 
In this case, the magnetization direction is fixed to the film plane. 
In other words, when $|H_{\rm appl}/H_{\rm s}^{(0)}| > 1$, the magnetization locates above the film plane, showing self-oscillation. 
Therefore, the self-oscillation is excited when current $I$ satisfies, 
\begin{equation}
  I
  \lesssim
  \frac{2 e MV}{\hbar \eta}
  H_{\rm appl},
  \label{eq:upper_limit}
\end{equation}
where we use the symbol $\lesssim$, because the approximation of $\lambda \to 0$ is used to derive Eq. (\ref{eq:upper_limit}). 
Equation (\ref{eq:upper_limit}) determines the saddle-node bifurcation of the spin-torque oscillator. 
We remind the reader that the applied field $H_{\rm appl}$ should satisfy 
\begin{equation}
  H_{\rm appl}
  >
  \frac{3 \lambda^{2}}{2-3 \lambda^{2}}
  \left(
    H_{\rm K}
    -
    4\pi M 
  \right),
  \label{eq:field_condition}
\end{equation}
to excite self-oscillation, as derived in Ref. [\onlinecite{taniguchi13}]. 



\section{Agility to current}
\label{sec:AppendixB}

The agility is defined as the frequency shift in response to an external force [\onlinecite{tamaru15}]. 
In the main text, we focused on the response of the spin-torque oscillator to the magnetic pulse 
because our work is focused on such experiments [\onlinecite{suto11,nagasawa11}]. 
Another interesting research target is the response to the current. 
For the spin-torque oscillator consisting of a perpendicularly magnetized free layer and an in-plane magnetized pinned layer, 
agility in response to the current is 
\begin{equation}
  \frac{\partial f}{\partial I}
  \simeq
  -\frac{\gamma}{2\pi}
  \frac{2 \alpha h}{(-2 \alpha + \lambda h_{\rm s})^{2}}
  \frac{\hbar \eta \lambda}{2eMV},
  \label{eq:agility_current}
\end{equation}
where we use Eqs. (\ref{eq:frequency}) and (\ref{eq:mz_stable}) 
In the experiments based on magnetic tunnel junctions, 
the agility in response to the voltage, rather than the current, might be useful [\onlinecite{tamaru15}]. 
Contrary to the agility in response to the magnetic field given by Eq. (\ref{eq:agility_field}), 
which is almost constant as $\gamma/(2\pi)$, 
Eq. (\ref{eq:agility_current}) varies for a wide range. 
For example, Eq. (\ref{eq:agility_current}) becomes $-[\gamma/(2\pi)] \times [\hbar \eta \lambda/(4 \alpha h eMV)]$ for the current $I/I_{\rm c} \simeq 1$, 
whereas it becomes zero for a large current limit $I \to \infty$.



\section{Critical exponent for thermally activated switching}
\label{sec:AppendixC}

In Sec. \ref{sec:Discussion}, we discuss the dynamical critical exponent of the phase transition 
between the critical point and the self-oscillation state of the spin-torque oscillator. 
A different type of critical exponent for the spin torque is discussed 
in a theoretical work considering the spin-torque switching of the magnetization in the thermally activated region [\onlinecite{taniguchi13PRB}]. 
Here, let us briefly discuss the relation between this past work and the current study. 
The spin-torque switching in the thermally activated region is described by the Fokker-Planck equation [\onlinecite{apalkov05}] 
\begin{equation}
  \frac{\partial \mathcal{P}}{\partial t}
  +
  \frac{\partial J}{\partial E}
  =
  0,
  \label{eq:Fokker_Planck}
\end{equation}
where $\mathcal{P}$ is the probability density of the magnetization distribution, 
whereas $J$ is the probability current density in the energy space given by 
\begin{equation}
  J
  =
  \mathscr{W}_{\alpha}
  \frac{d \mathscr{E}}{dE}
  \frac{\mathcal{P}}{\tau}
  +
  D 
  \frac{M}{\alpha\gamma}
  \mathscr{W}_{\alpha}
  \frac{\partial}{\partial E}
  \frac{\mathcal{P}}{\tau}.
  \label{eq:energy_current}
\end{equation}
Here, the effective energy density is defined as 
\begin{equation}
  \mathscr{E}(E)
  =
  \int^{E} 
  dE^{\prime} 
  \left[
    1
    +
    \frac{\mathscr{W}_{\rm s}(E^{\prime})}{\mathscr{W}_{\alpha}(E^{\prime})}
  \right].
\end{equation}
On the other hand, the second term in Eq. (\ref{eq:energy_current}) with the diffusion coefficient $D=\alpha \gamma k_{\rm B}T/(MV)$ 
represents the effect of the thermal fluctuation, 
where $k_{\rm B}$ is the Boltzmann constant and $V$ and $T$ are the volume and temperature of the free layer, respectively. 
The distribution function in a steady state is determined from Eq. (\ref{eq:energy_current}) as 
$\mathcal{P}/\tau \propto \exp[-\mathscr{E}V/(k_{\rm B}T)]$. 
The critical exponent $b$ of the spin-torque switching in the thermally activated region is defined as [\onlinecite{taniguchi13PRB}] 
\begin{equation}
  \int_{E_{\rm min}}^{E_{\rm saddle}}
  dE
  \left[
    1
    +
    \frac{\mathscr{W}_{\rm s}(E)}{\mathscr{W}_{\alpha}(E)}
  \right]
  =
  \Delta_{0}
  \left(
    1
    -
    \frac{I}{I_{\rm c}^{*}}
  \right)^{b}, 
  \label{eq:switching_barrier}
\end{equation}
where $E_{\rm saddle}$ is the saddle or maximum energy density of $E$, 
and $\Delta_{0}=(E_{\rm saddle}-E_{\rm min})V/(k_{\rm B}T)$ is the energy barrier separating the stable states of the free layer 
in the absence of the current. 
The scaling current $I_{\rm c}^{*}$ is defined as $\lim_{E \to E_{\rm saddle}}\mathscr{W}_{\rm s}/\mathscr{W}_{\alpha}=-I/I_{\rm c}^{*}$. 
In general, $I_{\rm c} \neq I_{\rm c}^{*}$, 
and $I^{*}/I_{\rm c}>1$ is a sufficient, but not a necessary, condition. 

The dynamical critical exponent $z$ in Eq. (\ref{eq:r_unstable}) is solely determined by the energy density corresponding to the critical point, 
whereas the exponent $b$ in Eq. (\ref{eq:switching_barrier}) is determined by the energy densities in the region of $[E_{\rm min},E_{\rm saddle}]$. 
For example, it is shown that the exponent $b$ in the in-plane magnetized system depends on the current magnitude, 
whereas that in the perpendicularly magnetized system is $2$ [\onlinecite{taniguchi13PRB}]. 



\section{A different definition of relaxation time}
\label{sec:AppendixD}

One might consider from Eq. (\ref{eq:LLG_energy}) that 
the relaxation time can be defined as 
\begin{equation}
  \int 
  dt 
  =
  \int_{E_{0}}^{E_{1}}
  \frac{d E}{\mathscr{W}_{\rm s}(E) + \mathscr{W}_{\alpha}(E)}, 
  \label{eq:relaxation_time_def_wrong}
\end{equation}
where $E_{0}$ and $E_{1}$ are energy densities corresponding to 
the constant energy curves at the initial and final states, respectively. 
However, Eq. (\ref{eq:relaxation_time_def_wrong}) is not suitable for defining the relaxation time. 
This is because $\mathscr{W}_{\rm s}(E_{1})+\mathscr{W}_{\alpha}(E_{1})=0$; 
therefore, the integrand diverges at the integral boundary, 
which would lead to the relaxation time derived from Eq. (\ref{eq:relaxation_time_def_wrong}) becoming infinite. 
This result can also be understood from Eq. (\ref{eq:E_solution_1}) that 
$\delta E$ decreases to zero with increasing time, but never becomes exactly zero. 
To avoid such divergence, Ref. [\onlinecite{yamada15}] for example replaces the integral boundary $E_{1}$ with a different value $E_{1}^{\prime}$ 
satisfying $E_{0} < E_{1}^{\prime} < E_{1}$ or $E_{1} < E_{1}^{\prime} < E_{0}$, 
and $\mathscr{W}_{\rm s}(E_{1}^{\prime})+\mathscr{W}_{\alpha}(E_{1}^{\prime}) \neq 0$. 
The value of $E_{1}^{\prime}$ in Ref. [\onlinecite{yamada15}] is determined 
from an assumption that the final state of the magnetization shifts from $E_{1}$ 
because of the thermal fluctuation, i.e., $|E_{1}-E_{1}^{\prime}| V \simeq k_{\rm B}T$. 
Then, a finite switching time can be obtained as in Ref. [\onlinecite{yamada15}]. 





\begin{thebibliography}{57}
\bibitem[{com()}]{comment1}
\bibinfo{note}{The coefficient $u^{\prime}$ in Eq. (\ref{eq:rate_u_prime}) is
  negative, but it is consistent with this argument. In Eq.
  (\ref{eq:LLG_approx_3}), we use $\delta m_{z}$, instead of $\delta E$. Since
  we focus on the critical slowing down near $m_{z}=+1$, $\delta m_{z}$ is
  negative, and therefore, $u^{\prime}$ should be negative to guarantee the
  monotonic relaxation of $\delta m_{z}$.}
\end{thebibliography}


\end{document}